\documentclass[letterpaper, 10 pt, conference]{ieeeconf}
\IEEEoverridecommandlockouts

\usepackage{cite}
\usepackage{float}
\usepackage{amsmath,amssymb,amsfonts}
\usepackage{algorithmic}
\usepackage{mathrsfs}
\usepackage{graphicx}
\usepackage{textcomp}
\usepackage{xcolor}
\usepackage[amsmath,thmmarks]{ntheorem}
\usepackage{hyperref}
\usepackage{enumerate}
\hypersetup{pdfborder={0 0 0},}
\def\BibTeX{{\rm B\kern-.05em{\sc i\kern-.025em b}\kern-.08em
    T\kern-.1667em\lower.7ex\hbox{E}\kern-.125emX}}
\usepackage{algorithm}

 \theoremheaderfont{\normalfont\bfseries} 
\theorembodyfont{\normalfont\itshape}    
\theoremseparator{.}                     

\newtheorem{remark}{Remark}
\newtheorem{theorem}{Theorem}
\newtheorem{lemma}{Lemma}

\newtheorem{assumption}{Assumption}
\newtheorem{proposition}{Proposition}

\begin{document}

\title{\LARGE \bf Distributed Optimization with Nesterov Momentum and Finite-Time Quantized Averaging over Directed Networks: Extended Version}

\author{Ruochen Wu, Xu Du, Mohammadreza Doostmohammadian, Karl~H.~Johansson, and Apostolos I. Rikos$^*$
\thanks{$^*$Corresponding author.}
	\thanks{Ruochen Wu, Xu Du, and Apostolos I. Rikos are with the Artificial Intelligence Thrust of the Information Hub, The Hong Kong University of Science and Technology (Guangzhou), Guangzhou, China. 
    Apostolos I. Rikos is also affiliated with the Department of Computer Science and Engineering, The Hong Kong University of Science and Technology, Hong Kong SAR, China. E-mails: {\tt~rwu856@connect.hkust-gz.edu.cn; \{michaelxudu, apostolosr\}@hkust-gz.edu.cn}. 
            }
           \thanks{Mohammadreza Doostmohammadian is with the Faculty of Mechanical Engineering, Semnan University, Semnan, Iran. E-mail: {\tt~doost@semnan.ac.ir}.
            } 
            \thanks{Karl H.~Johansson is with the Division of Decision and Control Systems, KTH Royal Institute of Technology, SE-100 44 Stockholm, Sweden. 
    He is also affiliated with Digital Futures, SE-100 44 Stockholm, Sweden. 
    E-mail:{\tt~kallej@kth.se}.
            }
            \thanks{The work of R.W., X.D., and A.I.R. was supported by the Guangzhou-HKUST(GZ) Joint Funding Scheme (Grant No. 2025A03J3960). The work of A.I.R. was also supported by the Guangdong Provincial Project (Grant No. 2024QN11G109).} 
}

\maketitle

\begin{abstract}
In modern large-scale networked systems, rapidly solving optimization problems while utilizing communication resources efficiently is critical for addressing complex tasks.  
However, distributed optimization over bandwidth-limited networks must cope with two obstacles: limited-bandwidth communication, and slow convergence when the objective is ill-conditioned. 
In this paper, we consider unconstrained distributed optimization over strongly connected directed networks, where nodes exchange quantized (i.e., compressed) messages. 
We propose QANM (Quantized Averaged Nesterov Momentum), an algorithm that combines Nesterov momentum with a finite-time quantized averaging protocol. 
At every iteration, QANM enables all nodes to reach exact agreement, without requiring doubly stochastic weight matrices or undirected communication.
For strongly convex and smooth objectives, we show that the objective gap of QANM decays at the accelerated rate $1-\kappa^{-1/2}$ (where $\kappa$ is the condition number), up to a residual of order $\Delta^2$ determined by the quantization level $\Delta$. 
This improves the condition-number dependence of the transient iteration complexity from $\kappa$ to $\sqrt{\kappa}$ compared with gradient descent that utilizes the same quantized averaging mechanism.   
Finally, we evaluate QANM on a distributed sensor fusion problem, showing that our algorithm exhibits the fastest initial error reduction compared with other methods from current literature.  
\end{abstract} 

\section{Introduction}\label{sec_intro}

The rapid growth of large-scale networked systems, including Internet of Things (IoT) devices \cite{intro1, wang2025fixed}, cloud computing infrastructures \cite{intro2}, and multi-vehicle systems \cite{ziming_rikos_JICV_2026_RL}, has made distributed optimization an increasingly important research area \cite{survey_distributed}. 
Unlike centralized approaches, which depend on a single processing unit and a central coordinator, distributed optimization offers advantages in scalability, robustness, and privacy preservation \cite{dist_structure9}. 
In this setting, each node performs local computation and exchanges information only with its neighbors, so that all nodes collectively converge to a globally optimal solution without a central coordinator.


Distributed optimization algorithms can broadly be classified into two categories according to their solution strategy: (i) primal methods, such as DGD \cite{DGD} and EXTRA \cite{extra}, and (ii) dual (primal-dual) methods, such as those based on ADMM \cite{admm} or ALADIN \cite{aladin-based}. 
As primal methods exhibit a simple structure, low per-iteration computation and communication overhead, and strong scalability \cite{survey_distributed}, they are the focus of this work.


Early distributed optimization algorithms \cite{DGD, dist_structure4_lihuaxie} typically rely on two assumptions: (i) nodes can exchange high-precision real-valued messages without communication constraints, and (ii) standard gradient-based updates converge at a satisfactory rate. 
In practice, neither assumption holds in general, and each gives rise to distinct challenges.
The first challenge is limited communication. 
Practical channels have finite bandwidth and precision, so messages must be quantized before transmission. 
The resulting quantization errors complicate the convergence analysis and, if not handled carefully, can prevent convergence to the optimum or even cause divergence \cite{Berahas_Wei_CDC_2019_Nested}. 
The second challenge is slow convergence on ill-conditioned objectives. 
When the objective curves much more steeply along some directions than others, its level sets form long, narrow valleys. 
The negative gradient points mostly across the valley rather than along it, and the step size must be small enough to remain stable along the steepest direction. 
As a result, gradient descent oscillates between the valley walls while making slow progress along its floor, producing a zigzag trajectory \cite{dict_opti16}. 
For a strongly convex and smooth objective, this slowdown is governed by the condition number \(\kappa\), the ratio of the largest to the smallest curvature, and the number of iterations required grows proportionally to \(\kappa\) \cite{intro_neste23}. 

To address the first challenge, several distributed optimization methods based on quantized message passing have been proposed \cite{rikos_consensus, rikos2023distributed_cdc, rikos2023distributed2, rikos2024distributed, quantizer30, quantizer31, quantizer32, quantizer33, quantizer34, quantizer35, quantizer36}. 
These methods analyze the errors induced by communication and, in most cases, guarantee convergence to a neighborhood of the optimal solution. 
However, except for \cite{rikos_consensus, rikos2023distributed2, rikos2023distributed_cdc, rikos2024distributed}, they rely on restrictive assumptions on the underlying network topology, such as undirected graphs, weight-balanced directed graphs, or doubly stochastic weight matrices. 
Such requirements are restrictive in practice: undirected communication is often unrealistic in wireless networks, where link strengths are asymmetric, and balancing a directed network (for generating a doubly-stochastic matrix) requires additional preprocessing \cite{Rikos_Hadjicostis_Themis_Balancing_2014}. 
Moreover, these methods build on gradient descent and therefore inherit its slow convergence on ill-conditioned objectives. 
The works in \cite{rikos_consensus, rikos2023distributed2, rikos2023distributed_cdc, rikos2024distributed} remove these topology restrictions by combining gradient descent with finite-time quantized averaging over general directed networks. However, they still rely on standard gradient updates. On ill-conditioned objectives such updates zigzag across narrow valleys of the cost function and make slow progress along them, so the number of iterations needed to reach a given accuracy grows linearly with the condition number, as \(\mathcal{O}(\kappa\log(1/\varepsilon))\). 

To address the second challenge, Nesterov's accelerated gradient method \cite{intro_neste23} (for centralized optimization) improves the convergence rate from \(\mathcal{O}(1/k)\) to \(\mathcal{O}(1/k^2)\) for convex problems and from \(\mathcal{O}\big((1-\kappa^{-1})^k\big)\) to \(\mathcal{O}\big((1-\kappa^{-1/2})^k\big)\) for strongly convex problems.
Several studies \cite{rw_neste30, rw_neste31, rw_neste32, rw_neste33, rw_neste34, rw_neste36, rw_neste21} bring this acceleration to distributed optimization.
However, they require a communication network modeled as an undirected graph (possibly with doubly stochastic weight matrices). 
The strongly convex method of \cite{rw_neste32}, for example, attains \(\mathcal{O}\big((1-C\kappa^{-5/7})^k\big)\), which is faster than gradient descent but slower than the aforementioned centralized rate, whereas \cite{rw_neste33} attains optimal rates by performing several real-valued gossip rounds per gradient evaluation. 
Other studies \cite{rw_neste35, rw_neste22, rw_neste23, rw_neste20, rw_neste38} extend Nesterov-type methods to directed graphs through row- or column-stochastic weights.
However, these aforementioned works assume that nodes exchange high-precision real-valued messages. 
This limits their use in bandwidth-limited networks, where messages must be quantized or compressed before transmission. 
The few exceptions, such as \cite{DOOSTMOHAMMADIAN2025106062, rw_neste37}, that consider momentum-based methods with imperfect communication (e.g., logarithmic quantization) consider undirected communication topologies, or weight-balanced directed networks. 
Furthermore, in all aforementioned works, agreement among nodes is reached only asymptotically, so at any finite iteration the nodes still hold different estimates.
This may be problematic in realistic scenarios in which exact agreement is necessary. 

The aforementioned limitations reveal a significant gap in the existing literature. 
Quantized methods over general unbalanced directed networks rely on standard gradient updates, whereas most accelerated methods assume real-valued communication and reach agreement only asymptotically. 
Introducing quantization into accelerated methods is itself challenging as they are known to accumulate errors from inexact gradients and may even diverge \cite{devolder2014first, devolder2013first}, and the quantization error would enter every iteration. 
To the best of the authors' knowledge, no existing algorithm combines all of the following features: (i) operation over directed networks without doubly stochastic or weight-balanced matrices, (ii) quantized communication, (iii) accelerated convergence, with a \(\sqrt{\kappa}\) condition-number dependence, and (iv) exact agreement among all nodes at every iteration. 
Each of these properties has been studied in isolation or in partial combination, and their simultaneous integration remains an open problem in existing literature. 
This raises the question: can Nesterov momentum be integrated into a finite-time quantized averaging framework so that the accelerated rate survives quantization over directed networks?

\textbf{Main Contributions.} 
In this paper, we aim to address the aforementioned gap in the literature by combining Nesterov momentum with a finite-time quantized average consensus protocol. 
The resulting algorithm, QANM, enables nodes to exchange quantized messages over unbalanced directed networks, and agree exactly at every iteration, while retaining an accelerated convergence rate. 
Our contributions are the following:
\\ \textbullet \ We introduce QANM, a distributed optimization algorithm that combines Nesterov momentum (to accelerate convergence and reduce zigzagging) with finite-time quantized average consensus protocol, and operates over strongly connected unbalanced directed networks (see Algorithm~\ref{alg: Nesterov Momentum}). 
Nodes communicate by exchanging quantized messages, and all nodes reach exact agreement at every iteration. 
\\ \textbullet \ We prove that the objective gap of QANM decays at the accelerated rate \(1-\kappa^{-1/2}\) up to an explicit residual of order \(\Delta^2\), improving the condition-number dependence of the transient iteration complexity from $\kappa$ for QAGD to $\sqrt{\kappa}$ (see Theorem~\ref{theo:convergence_nesterov}).  
\\ \textbullet \ On a distributed sensor fusion problem for multidimensional target parameter estimation, we show that QANM compares favorably against algorithms from the current literature exhibiting the fastest initial error reduction. 

Our analysis builds on the inexact Nesterov framework \cite{devolder2014first, devolder2013first}, which is known to accumulate errors and be prone to divergence. 
Because all nodes agree exactly at every iteration, the consensus-error coupling (that arises in distributed Nesterov analyses such as \cite{rw_neste32}) disappears. 
This leads to a clean estimating-sequence argument, and our rate contains no spectral-gap term. 
We then introduce a shifted recursion, with a \(\lambda/\tau\) shift in the auxiliary function \(\Psi\), which absorbs the additive quantization error at every iteration without letting it accumulate. 






\section{Notation and Background}


The symbols $\mathbb{R}$, $\mathbb{Q}$, and $\mathbb{Z}$ denote the sets of real, rational, and integer numbers, respectively, with $\mathbb{Q}_{>0}$ denoting the positive rationals. 
Vectors are denoted by bold lowercase letters
(e.g., \(\mathbf{x}\in\mathbb{R}^{p}\)) and matrices by
capital letters (e.g., \(A\)).
The transposes of \(A\) and \(\mathbf{x}\) are
\(A^\top\) and \(\mathbf{x}^\top\), and
\(\langle \mathbf{x}, \mathbf{y} \rangle
= \mathbf{x}^\top \mathbf{y}\) is the Euclidean inner product.
For $a\in\mathbb{R}$, $\lfloor a\rfloor$ and $\lceil a\rceil$ denote the greatest integer not exceeding $a$ and the smallest integer not less than $a$ (applied to a vector, they act componentwise). 
The symbol $\mathbb{I}$ is the identity matrix of appropriate dimension, and $\operatorname{diag}(a_1,\dots,a_n)$ is the diagonal matrix with $a_1,\dots,a_n$ on its main diagonal. 
The symbol $\|\cdot\|$ denotes the Euclidean norm and $\|\cdot\|_\infty$ the infinity norm. 
We write $x_i^{[k]}$ for the variable of node $v_i$ at optimization iteration $k$, and $p$ the dimension of the decision variable.

\subsection{Graph-Theoretic Notions}

The communication network is modeled as a directed graph (digraph) $\mathcal{G}=(\mathcal{V},\mathcal{E})$, with node set $\mathcal{V}=\{v_1,\dots,v_n\}$, where the number of nodes is $n=|\mathcal{V}|\ge 2$, and edge set $\mathcal{E} \subseteq \{(v_i, v_j) \in \mathcal{V} \times \mathcal{V} \ | \ i \neq j \}$, where the number of edges is \( m = |\mathcal{E}| \) (does not count self-loops). 
An edge $(v_i,v_j)\in\mathcal{E}$ indicates that node $v_i$ can receive information from node $v_j$, but not necessarily vice versa. 
Each node also has a virtual self-loop, which we do not include in $\mathcal{E}$. 
The set of in-neighbors of $v_i$ is $N_i^-=\{v_j\in\mathcal{V}\setminus\{v_i\} \mid (v_i,v_j)\in\mathcal{E}\}$, with in-degree $D_i^-=|N_i^-|$. 
The set of out-neighbors of $v_i$ is $N_i^+=\{v_j\in\mathcal{V}\setminus\{v_i\} \mid (v_j,v_i)\in\mathcal{E}\}$, with out-degree $D_i^+=|N_i^+|$. 
A directed path of length $t$ from $v_j$ to $v_i$ is a sequence of nodes $v_j=v_{j_0},v_{j_1},\dots,v_{j_t}=v_i$ such that $(v_{j_{\tau+1}},v_{j_\tau})\in\mathcal{E}$ for $\tau=0,\dots,t-1$. 
A digraph is strongly connected if there exists a directed path from every node \( v_i \) to every other node \( v_l \), for all pairs of nodes \( (v_i, v_l) \in \mathcal{V} \).
The diameter $D$ of a strongly connected digraph is the longest of the shortest paths between any two nodes \( v_i \), \( v_j \in \mathcal{V} \) in the network.

\subsection{Quantizers}\label{quantizers_subsec}

Quantizers are widely used to model communication constraints and imperfect information exchange in networked systems \cite{quantizer29}, and to improve communication efficiency by reducing the number of bits required to represent each transmitted message.  
Scalar quantizers process each signal component independently and are classified as uniform or non-uniform according to their quantization intervals, and as symmetric or asymmetric with respect to the origin. 
In this work, we use a uniform asymmetric (floor) scalar quantizer (applied component-wise to vector-valued signals), to enable communication-efficient information exchange and to model errors caused by finite computational and communication precision in distributed networks.
For a vector $x \in \mathbb{R}^p$ and a quantization level $\Delta\in\mathbb{Q}_{>0}$, the quantizer is defined as
\begin{equation}\label{quantizer_defn}
    q_{\Delta}(x)_i = \Delta \left\lfloor \frac{x_i}{\Delta} \right\rfloor, \quad i=1,\ldots,p.
\end{equation}
It satisfies $0\le x_i-q_\Delta(x)_i<\Delta$ for every $i$, and hence
\begin{equation}\label{quantizer_error}
    \|x - q_\Delta(x)\|_\infty<\Delta, \qquad \|x-q_\Delta(x)\|\le\sqrt{p}\,\Delta.
\end{equation} 

\section{Problem Formulation}

Let us consider a network modeled as a digraph $\mathcal{G}=(\mathcal{V},\mathcal{E})$, with the following setting:
\\ \noindent \textbullet \  Each node $v_i\in\mathcal{V}$ is associated with a local objective function $f_i:\mathbb{R}^p\to\mathbb{R}$, exclusively known to $v_i$.
\\ \noindent \textbullet \  Communication among nodes is subject to limited channel capacity, and only quantized values represented by rational numbers can be exchanged. 
\\ Our goal is to design a distributed algorithm that enables all nodes to collaboratively compute an approximate solution to the following optimization problem: 
\begin{subequations}\label{equ:problem_define}
\begin{align}
    & \min_{x_1,\dots,x_n\in\mathbb{R}^p} \quad F(x_1,\dots,x_n)\equiv\sum_{i=1}^{n}f_i(x_i), \label{eq:P1_obj}\\
    & \text{s.t.}~ x_i=x_j,\ \forall v_i,v_j\in\mathcal{V}, \label{eq:P1_cons1}\\
    & \text{nodes communicate with quantized values,} \label{eq:P1_cons3}
\end{align}
\end{subequations}
where $x^*$ is the optimal solution to the problem. 
Equation \eqref{eq:P1_obj} expresses the minimization of the global cost (defined as the sum of all local costs).
Equation \eqref{eq:P1_cons1} requires all nodes to agree on a common estimation of the optimal solution.
Finally, \eqref{eq:P1_cons3} models the communication constraint, where nodes may only exchange quantized values with their neighbors due to limited channel bandwidth. 

\subsection*{Operational Assumptions}

We now make the following assumptions that are required for our algorithm design and the theoretical analysis. 

\begin{assumption}\label{ass:1}
    The communication network is modeled as a \textit{strongly connected} digraph $\mathcal{G} = (\mathcal{V}, \mathcal{E})$. 
    Also, every node $v_i$ knows the diameter of the network $D$ (or an upper bound $D' \geq D$), and a common quantization level $\Delta$. 
\end{assumption}



\begin{assumption}\label{ass:2}
The local cost function $f_i$ of each node $v_i \in \mathcal{V}$ is closed, proper, $\mu$-strongly convex and $L$-smooth. 
Specifically, for each local cost function $f_i$, there exist constants $\mu>0$ and $L>0$ such that
\begin{equation}\label{eq: mu}
\begin{aligned}
    f_i(x_\alpha)+\nabla f_i(x_\alpha)^\top (x_\beta-x_\alpha) +\frac{\mu}{2}\|x_\beta-x_\alpha\|^2\leq f_i(x_\beta),
\end{aligned}
\end{equation}
and 
\begin{equation}\label{eq: lip}
\begin{aligned}
     \left\|\nabla f_i(x_\alpha)-\nabla f_i(x_\beta) \right\| \leq L \left\|x_\alpha-x_\beta\right\|, 
\end{aligned}
\end{equation}
for every $x_\alpha, x_\beta \in \mathbb R^p$. 
\end{assumption}




Assumption~\ref{ass:1} ensures information propagation among every node in the network (since there is a path from every node to every other node). 
Additionally, the network diameter (or an upper bound on it) is needed for nodes to determine in finite time whether the quantized coordination procedure has converged, so that all nodes hold identical states and \eqref{eq:P1_cons1} is fulfilled. 
The common quantization level enables nodes to communicate by exchanging quantized messages, so that \eqref{eq:P1_cons3} is fulfilled. 
Assumption~\ref{ass:2} ensures that the global optimization problem admits a unique optimal solution. 
In particular, \eqref{eq: mu} characterizes the strong convexity of each $f_i$, which guarantees the existence and uniqueness of the global optimal solution and facilitates our convergence analysis. 
Inequality \eqref{eq: lip} describes the smoothness of each function, (i.e., the gradients are Lipschitz continuous with constant $L$), which supports stable gradient-based updates. 
These properties imply that the function $f(x):=\frac{1}{n}\sum_{i=1}^n f_i(x)$ (whose minimizer coincides with the optimal solution $x^*$ of \eqref{equ:problem_define}), is also $\mu$-strongly convex and $L$-smooth.

\section{Distributed Optimization with Quantized Communication and Nesterov Momentum}\label{sec:nesterov_momentum}

\subsection{Distributed Optimization with Quantized Communication Accelerated by Nesterov Momentum}


In this section we propose a distributed optimization algorithm, described below as Algorithm~\ref{alg: Nesterov Momentum} (QANM).
Following \cite{rw_neste32}, we assume that every local cost function $f_i$ is $\mu$-strongly convex and $L$-smooth with common constants $\mu>0$ and $L\geq\mu$. 
If the local functions instead have heterogeneous constants $\mu_i$ and $L_i$, our results still hold by taking $\mu \le \min_i \mu_i$ and $L \ge \max_i L_i$, as noted in \cite[Remark~$3$]{rw_neste32}. 

\begin{algorithm}[t]
	\caption{QANM: Quantized Averaged Nesterov Momentum}
	\label{alg: Nesterov Momentum}
	\textbf{Input.} A strongly connected digraph $\mathcal{G} = (\mathcal{V}, \mathcal{E})$ with $n=|\mathcal{V}|$ nodes and $m=|\mathcal{E}|$ edges, a common digraph diameter $D$ (or an upper bound $D'$), quantization level $\Delta\in\mathbb{Q}_{>0}$, and curvature constants $0 < \mu \leq L$ available to all nodes. 
    Each node $v_i\in\mathcal{V}$ has an initial state $x_i^{[0]}$ and a local cost function $f_i$. 
    For every $v_i, v_j \in \mathcal{V}$, we have $x_i^{[0]} = x_j^{[0]}$. 
    Assumptions~\ref{ass:1} and~\ref{ass:2} hold.\\
	\textbf{Initialization.} Each node sets $x_i^{[-1]}=x_i^{[0]}$, $\kappa=L/\mu$, $\alpha=1/L$, and $\beta=(\sqrt{\kappa}-1)/(\sqrt{\kappa}+1)$.\\
	\textbf{Iteration.} For $k=0,1,2,\ldots$, each node $v_i\in\mathcal{V}$ performs the following steps:
	\begin{enumerate}
		\item Computes the look-ahead point
		\begin{equation}\label{eq: look-ahead position}
			s_i^{[k]}=x_i^{[k]}+\beta\bigl(x_i^{[k]}-x_i^{[k-1]}\bigr).
		\end{equation}
		\item Takes a local gradient step from the look-ahead point:
		\begin{equation}
			z_i^{[k+1]}=s_i^{[k]}-\alpha\nabla f_i\bigl(s_i^{[k]}\bigr).
		\end{equation}
		\item Applies Algorithm~\ref{alg:Consensus} componentwise to obtain
		\begin{equation}
		x_i^{[k+1]}=\text{Algorithm~\ref{alg:Consensus}}\bigl (q_{\Delta}(z_i^{[k+1]}), D, \Delta \bigr).
		\end{equation}
	\end{enumerate}
	\textbf{Output.} Each node $v_i\in\mathcal{V}$ computes $x_i^{[k]}$ as an estimate of the optimizer $x^*$ of problem~\eqref{equ:problem_define}.
\end{algorithm}


\begin{algorithm}[t]
	\caption{FTQAC: Finite-Time Quantized Average Consensus} 
	\textbf{Input.} $q_{\Delta}(z_i^{[k+1]}), \; D, \; \Delta$. \\
    \textbf{Initialization.} Each node $v_i \in \mathcal{V}:$
    \begin{enumerate} 
    \item Assigns probability $b_{li}$ to each out-neighbor $v_l \in N_i^+ \cup \{v_i\}$ as
    \begin{equation}
    b_{li}=\left\{
    \begin{aligned} 
       &\frac{1}{1+ D_i^+}, \quad && \text{if } l=i \text{ or } v_l\in N_i^+,\\
       &0, \quad && \text{if } l\neq i \text{ and } v_l\notin N_i^+;
    \end{aligned}
    \right.
    \end{equation}
    \item sets $\rho_i = q_{\Delta}(z_i^{[k+1]}) / \Delta$, $\tilde z_i = 2$, $\tilde y_i = 2\rho_i$. 
	\end{enumerate} 
	\textbf{Iteration.} For $\lambda = 1,2,\cdots$, each node $v_i \in \mathcal V$ does:
	\begin{enumerate}
	\item \textbf{if} $\lambda \bmod (D) = 1$, sets $M_i = \lceil \tilde y_i / \tilde z_i\rceil$ and $m_i = \lfloor \tilde y_i / \tilde z_i \rfloor$; 
	\item broadcasts its stopping variables $M_i, m_i \in \mathbb{Z}$ to every out-neighbor $v_l \in N_i^+$, and receives $M_j, m_j$ from every in-neighbor $v_j \in N_i^-$. 
    Then sets
	\[
	M_i=\max_{v_j\in N_i^- \cup \{v_i\}} M_j, \qquad
	m_i=\min_{v_j\in N_i^- \cup \{v_i\}} m_j;
	\]
	\item sets $\tilde c_i^z = \tilde z_i$;
	\item \textbf{while} $\tilde c_i^z > 1$ \textbf{do}
	\begin{enumerate}
	\item $\tilde c_i^y = \lfloor \tilde y_i / \tilde z_i \rfloor$;
	\item sets its mass variables as $\tilde y_i = \tilde y_i - \tilde c_i^y$, $\tilde z_i = \tilde z_i - 1$, and $\tilde c_i^z = \tilde c_i^z - 1$;
	\item transmits $\tilde c_i^y$ to a randomly chosen out-neighbor $v_l\in N_i^+ \cup \{v_i\}$ according to $b_{li}$;
	\item receives $\tilde c_j^y$ from $v_j \in N_i^- \cup \{v_i\}$ and sets
	\begin{equation}
	    \tilde y_i = \tilde y_i + \sum_{j=1}^{n} w_{\lambda,ij} \tilde c_j^y,
	\end{equation}
	\begin{equation}
	    \tilde z_i = \tilde z_i + \sum_{j=1}^{n} w_{\lambda,ij},
	\end{equation}
	where $w_{\lambda,ij} = 1$ when node $v_i$ receives the pair $\tilde c_j^y$, $1$, from node $v_j$ at time step $\lambda$ (otherwise $w_{\lambda,ij} = 0$ and $v_i$ receives no message at time step $\lambda$ from $v_j$);
	\end{enumerate}
	\item \textbf{if} $\lambda \bmod D=0$, \textbf{then, if} $ M_i-m_i\ \le 1$, then sets $x_i^{[k+1]}=m_i\Delta$ and stops the operation of the algorithm.
	\end{enumerate}
	\textbf{Output.} $x_i^{[k+1]}$.
	\label{alg:Consensus}
\end{algorithm}

The intuition of Algorithm~\ref{alg: Nesterov Momentum} (QANM) is as follows.  
Initially, each node $v_i$ maintains an estimate $x_i^{[0]}$ of the optimal solution. 
At each iteration $k$, each node computes a look-ahead point from its current and previous estimates. 
Then it takes a local gradient step from that point.
These two steps are applied locally at each node, and follow Nesterov's accelerated scheme for strongly convex functions \cite{intro_neste23, rw_neste32}.
Subsequently, each node applies the finite-time quantized average consensus protocol of Algorithm~\ref{alg:Consensus} (FTQAC), which gives all nodes the same quantized average of the local updates (a detailed description of the operation of Algorithm~\ref{alg:Consensus} (FTQAC) is shown in \cite{rikos_consensus_oscil_autom}). 
Hence, after every iteration, the nodes states are in exact agreement, and the averaged iterates follow Nesterov's centralized scheme perturbed, however, by the quantization error. 
It is important to note that the quantization level $\Delta$ used in FTQAC (i) is the same for every node, (ii) enables quantized communication among nodes, and (iii) determines the attainable precision of the solution.
FTQAC runs between two consecutive optimization iterations $k$ and $k+1$ of QANM, and for this reason it uses a different time index $\lambda$ rather than $k$. 
Furthermore, during initialization, all nodes have the same initial state (i.e., for every $v_i, v_j \in \mathcal{V}$, we have $x_i^{[0]} = x_j^{[0]}$). 
This can be achieved by executing a finite time consensus algorithm (e.g., a max-consensus or FTQAC), before we execute the main operations of QANM. 

\begin{remark}[Analysis Challenges and Future Directions]
It is important to note that analyzing QANM is challenging because accelerated methods are known to accumulate errors from inexact oracles and may even diverge \cite{devolder2014first, devolder2013first}.
In our QANM, the quantization error enters every iteration through the consensus step. 
However, in our analysis we show that this error does not accumulate, and the objective gap decays linearly up to a residual of order $\Delta^2$ (Theorem~\ref{theo:convergence_nesterov}).  
Additionally, a detailed analysis of the communication cost of FTQAC (e.g., the number of transmitted bits and the number of consensus rounds per iteration) is beyond the scope of this paper and is considered as future work.
\end{remark}

\subsection{Convergence of Algorithm~\ref{alg: Nesterov Momentum} (QANM)}
\label{sub:converge_analysis_without_4delta}

Before analyzing the convergence of QANM, we present the following proposition. 
It shows that the nodes agree exactly on their iterates after every iteration. 
Due to quantization, however, the common value is not necessarily equal to the real-valued average of the local updates. 
Its proof follows directly from \cite{rikos_consensus_oscil_autom, rikos_consensus} and is omitted. 

\begin{proposition}[Exact agreement of QANM iterates] \label{prop:exact_agreement} 
Under Assumption~\ref{ass:1}, FTQAC returns the same output at every node after each communication phase. 
Consequently, Step~3 of Algorithm~\ref{alg: Nesterov Momentum} guarantees that, for all $k \geq 1$ and any $v_i,v_j \in \mathcal{V}$, $ x_i^{[k]} = x_j^{[k]}$. 
\end{proposition}

We now analyze the convergence of QANM. 
Let us assume that $x_i^{[k]}\in \mathbb{R}^p$ for every node $v_i \in \mathcal{V}$. 
The iteration steps~$1$, $2$, and $3$ of QANM are: 
\begin{subequations}
    \begin{align}
        s_i^{[k]} &= x_i^{[k]} + \beta (x_i^{[k]} - x_i^{[k-1]}), 
        \label{equ:momentum_update} \\
        z_i^{[k+1]} &= s_i^{[k]} - \alpha \nabla f_i(s_i^{[k]}), 
        \label{equ:z_update} \\
        x_i^{[k+1]} &= \text{Algorithm \ref{alg:Consensus}}\big(q_{\Delta}(z_i^{[k+1]}),\;D,\;\Delta\big)
        \label{equ:consensus}
    \end{align}
\end{subequations}

Focusing on \eqref{equ:consensus}, for every node $v_i$ we have 
\begin{subequations}\small
\begin{align}
    x_i^{[k+1]} &= \frac{1}{n} \left( \sum_{j=1}^n \Delta \left\lfloor \frac{z_j^{[k+1]}}{\Delta} \right\rfloor \right) - \rho_i^{[k+1]}, \:\left\|\rho_i^{[k+1]}\right\|_\infty \leq \Delta, \\
    x_i^{[k]} &= \Delta \left\lfloor \frac{x_i^{[k]}}{\Delta} \right\rfloor + \epsilon_i^{[k]}, \:\left\|\epsilon_i^{[k]}\right\|_\infty \leq \Delta,
\end{align}\label{equ:consensus_error}
\end{subequations}
during each time step $k$, where $\rho_i^{[k]}$ and $\epsilon_i^{[k]}$ represent the cumulative error during the operation of Algorithm~\ref{alg:Consensus} (FTQAC), and the error due to quantization in \eqref{quantizer_defn}, respectively. 
We denote $f(x):=\frac{1}{n}\sum_i f_i(x)$, $\tau := \sqrt{\frac{\mu}{L}}$, and
 \begin{subequations}\label{equ:define_hat_zxs}
     \begin{align}
         \hat{z}^{[k+1]} & := \frac{1}{n} \sum_{i=1}^n z_i^{[k+1]}, \label{equ:define_hat_z} \\
         \hat{s}^{[k+1]} & := \frac{1}{n} \sum_{i=1}^n s_i^{[k+1]},  \label{equ:define_hat_s}
     \end{align}
\end{subequations}
for every $k \geq 0$. 
\begin{lemma}\label{lemma:lemma_1}
    For each node at time step $k \geq 1$, the following inequality holds:
    \begin{equation}\label{equ:lemma_1_main}
    \begin{split}
        \| x_i^{[k]} - \hat{z}^{[k]} \| & \leq 2\sqrt{p}\Delta. 
     \end{split}
\end{equation}
    \textbf{Proof.} See Appendix \ref{APP: 1}. \hfill $\blacksquare$
\end{lemma}
For convenience, we denote 
\begin{subequations}
    \begin{align}
        \lambda_i^{[k]} &:= x_i^{[k]} - \hat{z}^{[k]}, \\
        g^{[k]} &:= \nabla f(\hat{s}^{[k]}), \\
        \iota_i^{[k]}(x) &:= f(s_i^{[k]}) + \langle g^{[k]}, x-s_i^{[k]} \rangle + \frac{\mu}{2} \left\| x-s_i^{[k]} \right\|^2. \label{equ:define_iota} 
    \end{align}
\end{subequations} 
We introduce a sequence of auxiliary functions $\{\Psi_i^{[k]}(x)\}_{k \geq 1}$
satisfying the following recursive relation 
\begin{equation}
    \begin{aligned}
    \Psi_i^{[k+1]} (x) &= (1-\tau)\Psi_i^{[k]} \left(x- \frac{\lambda_i^{[k+2]}}{\tau} \right) \\
    & +\tau \iota_i^{[k+1]} \left(x- \frac{\lambda_i^{[k+2]}}{\tau} \right) + \frac{L}{2} \left\| \lambda_i^{[k+2]} \right\|^2, 
    \label{equ:Psi_iterate}
\end{aligned}
\end{equation}
with the initial function defined as  
\begin{align}
    \Psi_i^{[1]}(x) = f(x_i^{[2]}) + \frac{\mu}{2} \left\| x - \left[ s_i^{[2]} + \frac{1}{\tau}(s_i^{[2]} - x_i^{[2]}) \right] \right\|^2. \label{equ:Psi_initial} 
\end{align} 
We now denote the minimum value and the unique minimizer
of $\Psi_i^{[k]}(x)$, respectively, by
\begin{subequations}
    \begin{align}
        \overset{\ast}{\Psi}_i^{[k]} &:= \min_x \Psi_i^{[k]}(x), \\
        \phi_i^{[k]} &:= \arg\min_x \Psi_i^{[k]}(x). 
    \end{align}
\end{subequations}
Our analysis below follows the estimating-sequence technique for inexact Nesterov methods. 
We first construct a sequence of auxiliary quadratic functions and characterize their minimizers and minimum values (see Lemmas~\ref{lemma:lemma_21} and~\ref{lemma:lemma_22}).
Then, we bound these functions at the optimal solution and relate their minimum values to the objective values at the iterates (see Lemmas~\ref{lemma:lemma_23} and~\ref{lemma:lemma_24}). 
Finally, we combine these results to show accelerated convergence of the nodes' objective values up to a quantization-dependent neighborhood of the optimal solution (see Theorem~\ref{theo:convergence_nesterov}). 
\begin{lemma}\label{lemma:lemma_21}
    For every $k\geq 1$, $\Psi_i^{[k]}(x)$ is a convex quadratic function with a Hessian matrix equal to $\mu \mathbb{I}$, i.e., 
    \begin{align}
        \Psi_i^{[k]}(x) = \overset{\ast}{\Psi}_i^{[k]} + \frac{\mu}{2}\left\| x-\phi_i^{[k]} \right\|^2 , 
    \end{align}
    and $\{\phi_i^{[k]}\}$ and $\{ \overset{\ast}{\Psi}_i^{[k]} \}$ satisfy
    \begin{subequations}
    \begin{align}
        \phi_i^{[k+1]} &= (1-\tau)\phi_i^{[k]} + \tau s_i^{[k+1]} - \frac{\tau}{\mu}g^{[k+1]} + \frac{\lambda_i^{[k+2]}}{\tau}, \label{equ:lemma_21_phi_iterate}\\
        \overset{\ast}{\Psi}_i^{[k+1]} &= (1-\tau)\overset{\ast}{\Psi}_i^{[k]}  + \tau f(s_i^{[k+1]})- \frac{\tau^2}{2\mu}\left\| g^{[k+1]} \right\|^2 \nonumber\\ 
        &\quad+ (1-\tau)\tau \langle g^{[k+1]}, \phi_i^{[k]} - s_i^{[k+1]}\rangle + \frac{L}{2}\left\| \lambda_i^{[k+2]} \right\|^2 \nonumber\\ 
        &\quad + \frac{\mu(1-\tau)\tau}{2} \left\| \phi_i^{[k]} - s_i^{[k+1]} \right\|^2.
        \label{equ:lemma_21_star_Psi_iterate}
    \end{align}
    \end{subequations}
    \textbf{Proof.}
    See Appendix~\ref{APP: 21}. 
    \hfill $\blacksquare$
\end{lemma}


Before continuing our analysis, let us now note that for all nodes we have, \(x_i^{[0]}=x_j^{[0]}\) (and \(x_i^{[-1]}=x_i^{[0]}\) in Algorithm~1), and FTQAC returns the same output at every node (Proposition~1). 
Hence the cumulative quantization error \(\rho_i^{[k]}\) in \eqref{equ:consensus_error} is also the same for every node, and we have
\(
x_i^{[k]}=x_j^{[k]}\quad \text{for all } k\ge 0 \text{ and all } v_i,v_j\in\mathcal{V}.
\)
Therefore, the look-ahead points coincide as well, i.e., \(s_i^{[k]}=\hat{s}^{[k]}\) for every node \(v_i\) and every iteration \(k\), so that
\[
\nabla f\big(s_i^{[k]}\big)=\nabla f\big(\hat{s}^{[k]}\big)=\frac{1}{n}\sum_{j=1}^{n}\nabla f_j\big(s_j^{[k]}\big)=g^{[k]},
\]
and the averaged iterate satisfies \(\hat{z}^{[k+1]}=\hat{s}^{[k]}-\tfrac{1}{L}g^{[k]}\). 
In what follows let us also recall that \(\kappa=L/\mu\), \(\tau=\sqrt{\mu/L}=\kappa^{-1/2}\), and hence \(\beta=\tfrac{\sqrt{\kappa}-1}{\sqrt{\kappa}+1}=\tfrac{1-\tau}{1+\tau}\). 

\begin{lemma}\label{lemma:lemma_22}
    For every $k\geq1$, the sequence $\{ \phi_i^{[k]} \}$ satisfies the equations shown as below 
    \begin{align}
        \phi_i^{[k]} = \frac{1+\tau}{\tau}s_i^{[k+1]} - \frac{1}{\tau}x_i^{[k+1]}. \label{equ:lemma_22}
    \end{align}
    \textbf{Proof.}
    See Appendix~\ref{APP: 22}. 
    \hfill $\blacksquare$
\end{lemma}
\begin{lemma}\label{lemma:lemma_23}
    For every $k\geq1$, we have 
    \begin{equation}
        \begin{aligned}
        &\quad\quad\Psi_i^{[k+1]}\left( x+\frac{\lambda_i^{[k+2]}}{\tau} \right) - \frac{L}{2}\|\lambda_i^{[k+2]}\|^2 \\
        &\leq (1-\tau)\Psi_i^{[k]}(x) + \tau f(x). \label{equ:lemma_23_main}            
        \end{aligned}
    \end{equation}
    \textbf{Proof.}
    See Appendix~\ref{APP: 23}. 
    \hfill $\blacksquare$
\end{lemma}
\begin{lemma}\label{lemma:lemma_24}
    For every $k\geq1$, the sequence $\{\overset{*}{\Psi}_i^{[k]}\}$ satisfies 
    \begin{align}
        \overset{*}{\Psi}_i^{[k]}\geq f(x_i^{[k+1]}). \label{equ:lemma_24_main}
    \end{align}
    \textbf{Proof.}
    See Appendix~\ref{APP: 24}. 
    \hfill $\blacksquare$
\end{lemma}

\begin{theorem}\label{theo:convergence_nesterov}
    Suppose that Assumptions~\ref{ass:1} and \ref{ass:2} hold. Then, for every $k\geq 1$, as the condition number satisfies $\kappa>1$, the iterates $\{x_i^{[k]}\}_{i\in\mathcal{V}}$ generated by Algorithm~\ref{alg: Nesterov Momentum} (QANM) satisfy
    \begin{equation}
        \begin{aligned}
            f(x_i^{[k+1]}) - f(x^*) \leq 2(1-\tau)^{k-1} B_i^{[1]} + \mathcal{P}(\Delta^2), 
        \end{aligned}
    \end{equation}
    where $\Delta$ is the quantizer level and 
    \begin{subequations}
        \begin{align}
            &\Psi_i := f(x_i^{[2]}) + \frac{\mu}{2} \left\| x^* - \left[ s_i^{[2]} + \frac{1}{\tau}(s_i^{[2]} - x_i^{[2]}) \right] \right\|^2, \\
            &B_i^{[1]} := \Psi_i^{[1]}(x^*) - f(x^*), \\
            &\mathcal{P}(\Delta^2) := \frac{8Lp\Delta^2}{\left( 1-\sqrt{1-\tau} \right)^2}. 
        \end{align}
    \end{subequations}
    \textbf{Proof.} See Appendix~\ref{APP: 3}. 
    \hfill $\blacksquare$
\end{theorem}

\begin{remark}[Interpretation of Theorem~\ref{theo:convergence_nesterov}]
In Theorem~\ref{theo:convergence_nesterov} we show that fully distributed QANM with quantized communication achieves an accelerated linear convergence rate. 
With $\tau=\sqrt{\mu/L}$, the transient part of the objective gap decays geometrically at rate $1-\sqrt{\mu/L}=1-\kappa^{-1/2}$, while quantization contributes a residual of order $\Delta^2$. 
Hence, quantized communication among nodes limits our attainable accuracy but does not slow down the accelerated decay of the transient term. 
\end{remark}

\begin{remark}[Acceleration of QANM]\label{rem:acceleration_guarantee}
For a target transient error $\varepsilon$ above the quantization residual, Theorem~\ref{theo:convergence_nesterov} gives an iteration complexity of $K_{\mathrm{QANM}}=\mathcal{O}(\sqrt{\kappa}\log(1/\varepsilon))$. 
Under matching step-size and curvature settings, \cite[Theorem~1]{rikos_consensus} gives $K_{\mathrm{QAGD}}=\mathcal{O}(\kappa\log(1/\varepsilon))$ for a fixed network size. 
Thus, QANM improves the condition-number dependence from $\kappa$ to $\sqrt{\kappa}$, which is most relevant when $\kappa\gg1$. 
Since $1-\sqrt{1-\tau}\geq\tau/2$, the residual satisfies $\mathcal{P}(\Delta^2)\leq 32\,\kappa Lp\,\Delta^2$, which grows linearly with $\kappa$.
Initial-error constants inside the logarithms are omitted.
\end{remark}

\begin{remark}[Comparison with Current Literature]
For strongly convex and smooth objectives, \cite{rw_neste32} establishes the rate \(\mathcal{O}\big((1-C\kappa^{-5/7})^k\big)\), where \(C>0\) is independent of \(\kappa\) and depends on the second largest singular value of the mixing matrix. 
In contrast, Theorem~\ref{theo:convergence_nesterov} guarantees that QANM converges at the rate \(\mathcal{O}\big((1-\kappa^{-1/2})^k\big)\) to a neighborhood of the optimal function value when \(\tau=\sqrt{\mu/L}\). 
The corresponding iteration complexities are \(\mathcal{O}(\sqrt{\kappa}\log(1/\varepsilon))\) for QANM, for \(\varepsilon\) above the quantization residual, and \(\mathcal{O}(C^{-1}\kappa^{5/7}\log(1/\varepsilon))\) for Acc-DNGD-SC. 
Therefore, QANM has the better dependence on \(\kappa\) than Acc-DNGD-SC.  
APD-SC \cite{rw_neste22} and MSDA \cite{rw_neste33} match the \(\sqrt{\kappa}\) dependence. 
However, the advantage of QANM lies in scenarios it can handle. 
First, MSDA requires an undirected graph with a symmetric gossip matrix, Acc-DNGD-SC requires an undirected graph with doubly stochastic weights, and APD-SC operates over directed graphs requiring column-stochastic weights. 
Our QANM operates over general strongly connected directed networks without doubly stochastic or weight-balanced matrices. 
Second, during Acc-DNGD-SC, APD-SC, MSDA nodes exchange real-valued messages, whereas QANM enables nodes to communicate with quantized messages (and thus operate over bandwidth-limited networks). 
Third, QANM provides exact agreement at every iteration through FTQAC, whereas aforementioned algorithms reach agreement only asymptotically. 
Finally, the rate in Theorem~\ref{theo:convergence_nesterov} has no mixing-matrix constant, unlike the rates of Acc-DNGD-SC, APD-SC and MSDA, which depend on the spectral properties or eigengap of their matrices (note that in QANM, the network enters through the number of FTQAC rounds per iteration, whose analysis is left for future work). 
On the other hand, the real-valued methods converge to the exact optimum, whereas QANM converges to a neighborhood whose size is controlled by \(\Delta\), due to the nodes exchanging quantized-valued messages.
\end{remark}

\section{Application: Distributed Sensor Fusion for Multi-Dimensional Target Parameter Estimation}\label{sec:simulation}

\begin{figure}[t]
    \centering
    \includegraphics[width=0.47\textwidth]{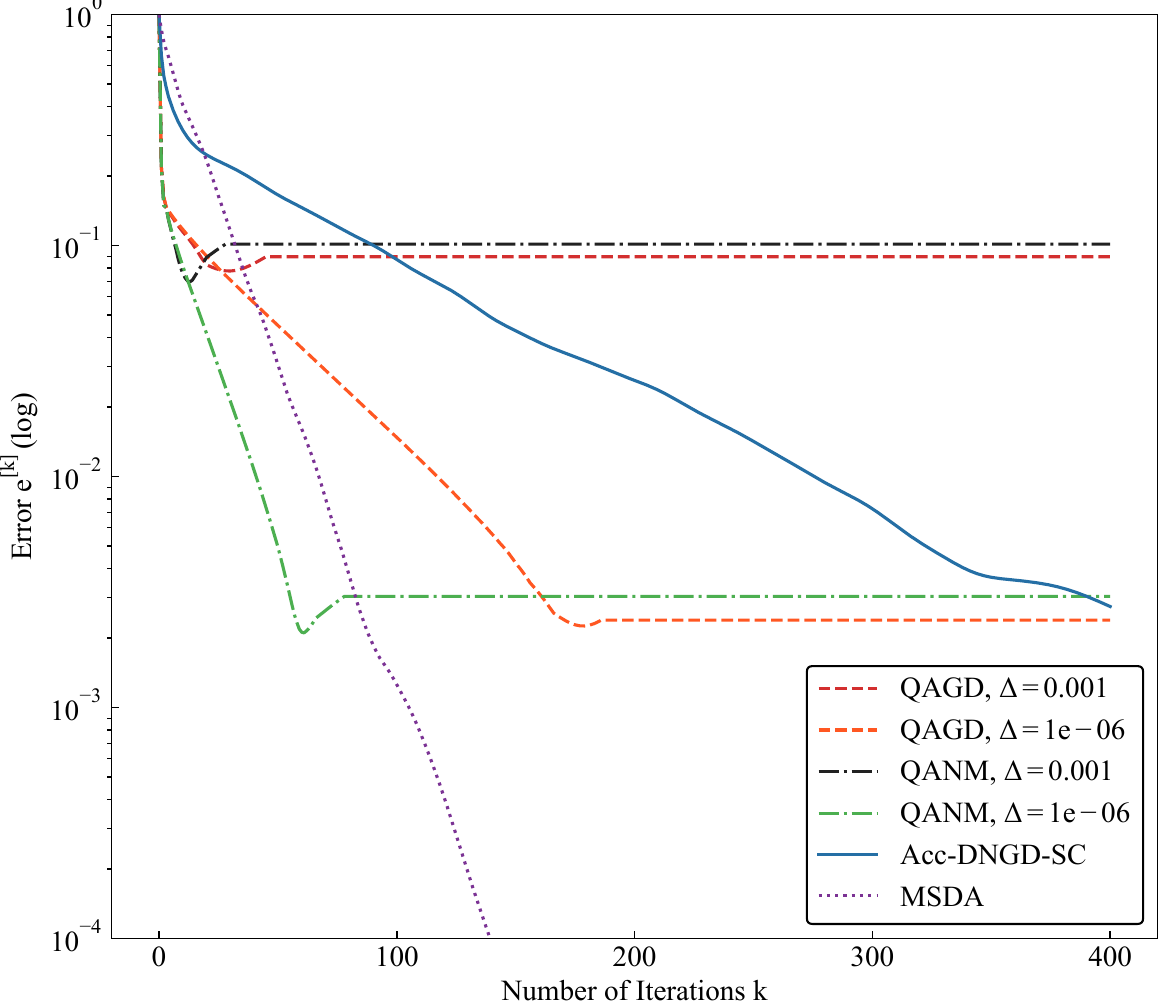}
    \caption{Normalized error $e^{[k]}$ (log scale) versus iterations $k$ for QANM, QAGD in \cite{rikos_consensus}, Acc-DNGD-SC in \cite{rw_neste32}, and MSDA in \cite{rw_neste33}, on the sensor fusion problem with $n=20$ nodes
    \protect}
    \label{fig:output_ideal}
\end{figure}

We evaluate Algorithm~\ref{alg: Nesterov Momentum} (QANM) on a distributed sensor fusion problem for multi-dimensional target parameter estimation \cite{simulate28}. 
A network of $n=20$ sensor groups (nodes) seeks to estimate a common target parameter from local measurements in dimension $p=5$. 
Each node $v_i$ holds a measurement vector $x_{0,i} \in \mathbb{R}^5$, whose components are drawn independently and uniformly from $\{1,2,3,4,5\}$, and a weight $\omega_i$ drawn uniformly from $\{1,2,3,4,5\}$. 
The local cost function is 
\begin{equation}\label{equ:simulation_xpx}
f_i(x)=\frac{\omega_i}{2}(x-x_{0,i})^\top P\,(x-x_{0,i}),
\end{equation}
where $P=\mathrm{diag}\big(\tfrac{1}{16},\tfrac{1}{8},\tfrac{1}{4},\tfrac{1}{2},1\big)$ is shared by all nodes, and the nodes seek $x^*=\arg\min_{x}\sum_{i=1}^{n}f_i(x)$. 
The matrix $P$ makes the measurement errors have different impact across dimensions, which produces the ill-conditioning (zigzag behavior) that momentum is meant to mitigate.
We compare QANM with QAGD \cite[Algorithm~1]{rikos_consensus}, Acc-DNGD-SC \cite{rw_neste32}, and MSDA \cite[Algorithm~2]{rw_neste33}. 
QANM and QAGD use a randomly generated strongly connected digraph with the same underlying $4 \times 5$ grid topology, and exchange quantized messages with quantization level $\Delta \in \{10^{-3}, 10^{-6}\}$. 
Acc-DNGD-SC and MSDA use the bidirectional $4\times5$ grid with real-valued communication among nodes. 
To compare the methods, we use the normalized error
\begin{equation}\label{equ:define_error}
e^{[k]}=\left(\frac{1}{n}\sum_{i=1}^{n}\frac{\|x_i^{[k]}-x^*\|}{\|x_i^{[0]}-x^*\|}\right)^{1/2}. 
\end{equation}
For each method, we test $1000$ candidate step sizes and select the one for which $e^{[k]}<10^{-2}$ is reached earliest and holds for all iterations up to $k=2000$. 
Figure~\ref{fig:output_ideal} plots $e^{[k]}$ against the iteration $k$ on a logarithmic scale for both quantization levels.


As shown in Fig.~\ref{fig:output_ideal}, QANM and QAGD reduce the error rapidly before reaching quantization-dependent error floors, which decrease with \(\Delta\). 
QANM decreases faster than QAGD at both quantization levels. 
For \(\Delta=10^{-6}\), QANM reaches and remains below \(e^{[k]}=10^{-2}\) after $41$ iterations, compared with $117$ for QAGD, $68$ for MSDA, and $276$ for Acc-DNGD-SC. 
Thus, QANM shows the fastest initial error reduction among the four methods in the tested setting. 
Note however that the real-valued benchmarks have no error floor. 
Specifically, MSDA continues to decrease and passes below the quantized floors, whereas Acc-DNGD-SC decreases at a slower rate. 
These differences follow from how the algorithms operate and their requirements regarding the underlying network (as identified in Section~\ref{sec_intro}). 
QAGD in \cite{rikos_consensus} uses the same finite-time quantized averaging (FTQAC) as QANM, with quantized messages, directed networks and exact agreement, but without momentum. 
Therefore the gap between them QAGD and QANM isolates the effect of Nesterov acceleration. 
The algorithms Acc-DNGD-SC \cite{rw_neste32} and MSDA \cite{rw_neste33} are accelerated but require undirected graphs (also doubly-stochastic for Acc-DNGD-SC \cite{rw_neste32}) and nodes exchange real-valued messages. 
Acc-DNGD-SC uses one communication round per iteration, so agreement among nodes is only asymptotic, whereas MSDA is a dual method that uses multiple gossip rounds per gradient evaluation. 
QANM, in contrast, operates over strongly connected directed networks without requiring doubly stochastic weights, combining acceleration with quantized messages, and exhibiting exact agreement at every iteration.  

\section{Conclusions and Future Directions}

We introduced QANM, a distributed optimization algorithm that combines Nesterov momentum with finite-time quantized averaging (FTQAC) over strongly connected directed networks. 
During its operation nodes exchange quantized-valued messages, without requiring networks that are undirected and/or modeled as doubly stochastic matrices. 
We showed that QANM converges at an accelerated linear rate $1-\kappa^{-1/2}$, to a neighborhood of the optimal solution whose size is determined by the quantization level. 
In a sensor-fusion experiment, QANM showed the fastest initial error reduction among the tested methods. 

As future work we will analyze QANM with node-dependent momentum coefficients under quantized communication and characterize the coefficient ranges that ensure convergence. 
Additionally, we will analyze the communication cost and resource usage of QANM, and quantify the tradeoff between faster outer-iteration convergence and the total communication cost our protocol.




\appendices
\section{Proof of Lemma \ref{lemma:lemma_1}}\label{APP: 1}
For convenience, we denote
\begin{align}
    \bar{\epsilon}^{[k]} := \frac{1}{n}  \sum_{j=1}^n \left(\Delta \left\lfloor \frac{z_j^{[k]}}{\Delta} \right\rfloor - z_j^{[k]} \right), \nonumber
\end{align}
and from \eqref{equ:consensus_error}, we have $\|\bar{\epsilon}^{[k]}\|_\infty\leq \Delta$. 
Combining \eqref{equ:consensus_error} and \eqref{equ:define_hat_z}, we have
    \begin{equation}
        \begin{aligned}
            \left\| x_i^{[k]} - \hat{z}^{[k]} \right\| &= \left\| \frac{1}{n}  \sum_{j=1}^n \left(\Delta \left\lfloor \frac{z_j^{[k]}}{\Delta} \right\rfloor - z_j^{[k]} \right) - \rho_i^{[k]}\right\| \\
            &\leq \left\| \rho_i^{[k]}\right\| + \left\| \bar{\epsilon}^{[k]} \right\| \\
            & \leq 2\sqrt{p}\Delta.
        \end{aligned}
    \end{equation}
So Lemma~\ref{lemma:lemma_1} is proved. 

\section{Proof of Lemma \ref{lemma:lemma_21}}\label{APP: 21} 
    We prove by induction on $k$. \\
    \textbf{Base case:}~ For $k=1$, we have 
    \begin{align}
        \nabla^2\Psi_i^{[1]}(x) \overset{\eqref{equ:Psi_initial}}{=} \mu \mathbb{I}. 
    \end{align}
    \textbf{Induction hypothesis:}~
    Suppose that $P_k$ holds for some $k\geq1$, i.e.,
    \begin{align}
        \nabla^2 \Psi_i^{[k]}(x)
        = \mu\mathbb{I}.
        \label{equ:lemma_21_induct_hypoth}
    \end{align}
    \textbf{Inductive step:}~ From \eqref{equ:Psi_iterate}, we have 
    \begin{equation}
        \begin{aligned}
            \nabla^2 \Psi_i^{[k+1]}(x) &\overset{\eqref{equ:Psi_iterate}}{=} (1-\tau)\nabla^2 \Psi_i^{[k]}\left(x-\frac{\lambda_i^{[k+2]}}{\tau}\right) \\&\quad\quad + \tau \nabla^2 \iota_i^{[k+1]}\left(x-\frac{\lambda_i^{[k+2]}}{\tau}\right)= \mu \mathbb{I}. 
            \label{equ:lemma_21_induct_step}
        \end{aligned}
    \end{equation}
    Therefore, by the principle of mathematical induction, $P_k$ holds for every integer $k\geq1$. 
    Hence, $\Psi_i^{[k+1]}(x)$ is a quadratic function with Hessian $\mu\mathbb{I}$ and can be written as
    \begin{align}
        \Psi_i^{[k+1]}(x)
        =
        \overset{\ast}{\Psi}_i^{[k+1]}
        +
        \frac{\mu}{2}
        \left\|x-\phi_i^{[k+1]}\right\|^2. \label{equ:lemma_21_quadra_express}
    \end{align}
    For convenience, for $k\geq1$, we denote 
    \begin{subequations}
    \begin{align}
        y&:=x-\frac{\lambda_i^{[k+2]}}{\tau}, \label{equ:define_y}\\
        u_i^{[k+1]} &:= (1-\tau)\phi_i^{[k]} - \frac{\tau}{\mu}g^{[k+1]} + \tau s_i^{[k+1]}. \label{equ:define_u}
    \end{align}
    \end{subequations}
    Combining \eqref{equ:lemma_21_quadra_express} with \eqref{equ:Psi_iterate}, we have
    \begin{equation}
        \begin{aligned}
            &\quad\quad \Psi_i^{[k+1]}(x) \\
            &\overset{\eqref{equ:lemma_21_quadra_express}}{=} (1-\tau) \left( \overset{*}{\Psi}_i^{[k]} + \frac{\mu}{2}\left\| y-\phi_i^{[k]} \right\|^2 \right) +\tau \iota_i^{[k+1]} \left(y \right) \\
            &\quad + \frac{L}{2} \| \lambda_i^{[k+2]} \|^2 \\
            &\overset{\eqref{equ:define_iota}}{=} (1-\tau)\overset{*}{\Psi}_i^{[k]} + \frac{\mu}{2}\|y\|^2  + \langle y, -(1-\tau)\mu\phi_i^{[k]} \\
            &\quad + \tau g^{[k+1]} - \mu\tau s_i^{[k+1]}\rangle + \frac{(1-\tau)\mu}{2}\|\phi_i^{[k]}\|^2 \\
            &\quad + \tau f(s_i^{[k+1]}) - \tau\langle g^{[k+1]}, s_i^{[k+1]}\rangle + \frac{\mu\tau}{2}\|s_i^{[k+1]}\|^2 \\
            &\quad + \frac{L}{2} \| \lambda_i^{[k+2]} \|^2 \\
            &\overset{\eqref{equ:define_u}}{=}(1-\tau)\overset{*}{\Psi}_i^{[k]} + \frac{\mu}{2}\left\| y - u_i^{[k+1]} \right\|^2\\
            &\quad + \frac{(1-\tau)\mu}{2}\|\phi_i^{[k]}\|^2 + \tau f(s_i^{[k+1]}) + \frac{\mu\tau}{2}\|s_i^{[k+1]}\|^2 \\
            &\quad - \tau\langle g^{[k+1]}, s_i^{[k+1]}\rangle + \frac{L}{2} \|\lambda_i^{[k+2]}\|^2 - \frac{\mu}{2}\|u_i^{[k+1]}\|^2 \\
            &= (1-\tau)\overset{*}{\Psi}_i^{[k]} + \frac{\mu}{2}\left\| x - \phi_i^{[k+1]} \right\|^2 + \tau f(s_i^{[k+1]}) \\
            &\quad + \frac{\mu\tau(1-\tau)}{2}\left\| \phi_i^{[k]} - s_i^{[k+1]}\right\|^2 + \tau(1-\tau) \langle g^{[k+1]}\\
            &\quad , \phi_i^{[k]} - s_i^{[k+1]}\rangle - \frac{1}{2L}\|g^{[k+1]}\|^2 + \frac{L}{2}\|\lambda_i^{[k+2]}\|^2. 
            \label{equ:lemma_21_main_progress}
        \end{aligned}
    \end{equation}
    So from \eqref{equ:lemma_21_main_progress}, we can get \eqref{equ:lemma_21_phi_iterate} and \eqref{equ:lemma_21_star_Psi_iterate}. 
    Thus, Lemma~\ref{lemma:lemma_21} is then proved.

\section{Proof of Lemma~\ref{lemma:lemma_22}} \label{APP: 22}
    We prove by induction on $k$. \\
    \textbf{Base case:}~For $k=1$, we have
    \begin{align}
        \phi_i^{[1]} \overset{\eqref{equ:Psi_initial}}{=} s_i^{[2]} + \frac{1}{\tau}(s_i^{[2]} - x_i^{[2]}) = \frac{1+\tau}{\tau}s_i^{[2]} - \frac{1}{\tau}x_i^{[2]}. 
    \end{align}
    \textbf{Induction hypothesis:}~Suppose that $P_k$ holds for some $k\geq1$, i.e., 
    \begin{align}
        \phi_i^{[k]} = \frac{1+\tau}{\tau}s_i^{[k+1]} - \frac{1}{\tau}x_i^{[k+1]}. \label{equ:lemma_22_induct_hypo}
    \end{align}
    \textbf{Induction step:}~From Step~1, we can get
    \begin{equation}
        \begin{aligned}
            &\quad\quad (1+\tau)s_i^{[k+2]} \\
            &\overset{\eqref{eq: look-ahead position}}{=} (1+\tau)x_i^{[k+2]} + (1-\tau)\left(x_i^{[k+2]} - x_i^{[k+1]} \right) \\
            &=2x_i^{[k+2]} - (1-\tau)x_i^{[k+1]}\\
            &\overset{\eqref{equ:z_update}}{=} 2\left( s_i^{[k+1]}-\frac{g^{[k+1]}}{L} + \lambda_i^{[k+2]} \right)- (1-\tau)x_i^{[k+1]}\\
            &\overset{\eqref{equ:lemma_22_induct_hypo}}{=} 2\left( s_i^{[k+1]}-\frac{g^{[k+1]}}{L} + \lambda_i^{[k+2]} \right)- (1-\tau^2)s_i^{[k+1]}\\
            &\quad + (1-\tau)\tau\phi_i^{[k]}\\
            &= (1+\tau^2)s_i^{[k+1]} + (1-\tau)\tau\phi_i^{[k]} - \frac{2}{L}g^{[k+1]} + 2\lambda_i^{[k+2]}. 
            \label{equ:lemma_22_progress_1}
        \end{aligned}
    \end{equation}
    From \eqref{equ:lemma_21_phi_iterate}, we have
    \begin{equation}
        \begin{aligned}
            &\quad\quad x_i^{[k+2]} + \tau\phi_i^{[k+1]} \\
            &\overset{\eqref{equ:lemma_21_phi_iterate}}{=}
            \left(s_i^{[k+1]} - \frac{g^{[k+1]}}{L} + \lambda_i^{[k+2]} \right) + (1-\tau)\tau \phi_i^{[k]}  \\
            &\quad+ \tau^2s_i^{[k+1]} - \frac{\tau^2}{\mu}g^{[k+1]}\\
            &= (1+\tau^2)s_i^{[k+1]} + (1-\tau)\tau\phi_i^{[k]} - \frac{2}{L}g^{[k+1]} + 2\lambda_i^{[k+2]}. \label{equ:lemma_22_progress_2}
        \end{aligned}
    \end{equation}
    Hence, combining \eqref{equ:lemma_22_progress_1} and \eqref{equ:lemma_22_progress_2}, we have
    \begin{align}
        (1+\tau)s_i^{[k+2]} = x_i^{[k+2]} + \tau\phi_i^{[k+1]}. 
    \end{align}
    Therefore, by the principle of mathematical induction, $P_k$ holds for every integer $k\geq1$. \\
    Hence, Lemma~\ref{lemma:lemma_22} is proved.

\section{Proof of Lemma~\ref{lemma:lemma_23}} \label{APP: 23}
    Substituting $x=x+\frac{\lambda_i^{[k+2]}}{\tau}$ into \eqref{equ:Psi_iterate}, we obtain
    \begin{equation}
        \begin{aligned}
            &\quad\quad\Psi_i^{[k+1]}\left( x+\frac{\lambda_i^{[k+2]}}{\tau} \right) - \frac{L}{2}\|\lambda_i^{[k+2]}\|^2 \\
            &\overset{\eqref{equ:Psi_iterate}}{=} (1-\tau)\Psi_i^{[k]}(x) + \tau\iota_i^{[k+1]}(x) \\
            &\overset{\eqref{equ:define_iota}}{\leq} (1-\tau)\Psi_i^{[k]}(x) + \tau f(x). 
        \end{aligned}
    \end{equation}
    Hence, Lemma~\ref{lemma:lemma_23} is proved.

\section{Proof of Lemma~\ref{lemma:lemma_24}} \label{APP: 24}
    We prove by induction on $k$. \\
    \textbf{Base case:}~For $k=1$, from \eqref{equ:Psi_initial}, we have
    \begin{align}
        \overset{*}{\Psi}_i^{[1]} \overset{\eqref{equ:Psi_initial}}{=} f(x_i^{[2]}). 
    \end{align}
    \textbf{Induction hypothesis:}~Suppose that $P_k$ holds for some $k\geq1$, i.e., 
    \begin{align}
        \overset{*}{\Psi}_i^{[k]}\geq f(x_i^{[k+1]}). \label{equ:lemma_24_induct_hypo}
    \end{align}
    \textbf{Induction step:}~From Lemma~\ref{lemma:lemma_22}, we have
    \begin{align}
        \tau(\phi_i^{[k]} - s_i^{[k+1]}) \overset{\eqref{equ:lemma_22}}{=} s_i^{[k+1]} - x_i^{[k+1]}. \label{equ:lemma_24_formula_B}
    \end{align}
    For convenience, we denote
    \begin{align}
        \tilde{\Psi}_i^{[k+1]} := \overset{*}{\Psi}_i^{[k+1]} - \frac{L}{2}\|\lambda_i^{[k+2]}\|^2. \label{equ:define_tilde_Psi}
    \end{align}
    From Lemma~\ref{lemma:lemma_21}, we have
    \begin{align}
        \tilde{\Psi}_i^{[k+1]} &\overset{\eqref{equ:lemma_21_star_Psi_iterate}}{=} (1-\tau)\overset{\ast}{\Psi}_i^{[k]}  + \tau f(s_i^{[k+1]})- \frac{\tau^2}{2\mu}\left\| g^{[k+1]} \right\|^2 \nonumber\\ 
        &\quad+ (1-\tau)\tau \langle g^{[k+1]}, \phi_i^{[k]} - s_i^{[k+1]}\rangle  \nonumber\\ 
        &\quad + \frac{\mu(1-\tau)\tau}{2} \left\| \phi_i^{[k]} - s_i^{[k+1]} \right\|^2.
        \label{equ:lemma_24_formula_C}
    \end{align}
    Substituting $x_\alpha=s_i^{[k+1]}, x_\beta = x_i^{[k+1]}$ into \eqref{eq: mu}, we obtain
    \begin{equation}
        \begin{aligned}
            \overset{*}{\Psi}_i^{[k]} &\overset{\eqref{equ:lemma_24_induct_hypo}}{\geq} f(x_i^{[k+1]})\\
            &\overset{\eqref{eq: mu}}{\geq} f(s_i^{[k+1]})
             + \langle g^{[k+1]}, x_i^{[k+1]}-s_i^{[k+1]}\rangle \\
             &\quad +\frac{\mu}{2}\|x_i^{[k+1]}-s_i^{[k+1]}\|^2\\
             &\overset{\eqref{equ:lemma_24_formula_B}}{=} f(s_i^{[k+1]}) - \tau\langle g^{[k+1]}, \phi_i^{[k]} - s_i^{[k+1]} \rangle\\
             &\quad + \frac{\mu\tau^2}{2}\left\| \phi_i^{[k]} - s_i^{[k+1]} \right\|^2. 
             \label{equ:lemma_24_progress_1}
        \end{aligned}
    \end{equation}
    Combining \eqref{equ:lemma_24_formula_C} with \eqref{equ:lemma_24_progress_1}, we can obtain
    \begin{equation}
        \begin{aligned}
            \overset{*}{\Psi}_i^{[k+1]} &\overset{\eqref{equ:define_tilde_Psi}}{=} \tilde{\Psi}_i^{[k+1]} + \frac{L}{2}\|\lambda_i^{[k+2]}\|^2 \\
            &\overset{\eqref{equ:lemma_24_formula_C}}{=} (1-\tau)\overset{\ast}{\Psi}_i^{[k]}  + \tau f(s_i^{[k+1]})- \frac{\tau^2}{2\mu}\left\| g^{[k+1]} \right\|^2 \nonumber\\ 
            &\quad+ (1-\tau)\tau \langle g^{[k+1]}, \phi_i^{[k]} - s_i^{[k+1]}\rangle  \nonumber\\ 
            &\quad + \frac{\mu(1-\tau)\tau}{2} \left\| \phi_i^{[k]} - s_i^{[k+1]} \right\|^2 + \frac{L}{2}\|\lambda_i^{[k+2]}\|^2 \\
            &\overset{\eqref{equ:lemma_24_progress_1}}{\geq} f(s_i^{[k+1]}) - \frac{\tau^2}{2\mu}\left\| g^{[k+1]} \right\|^2 + \frac{L}{2}\|\lambda_i^{[k+2]}\|^2\\
            &\quad + \frac{\mu(1-\tau^2)\tau}{2}\|\phi_i^{[k]} - s_i^{[k+1]}\|^2 \\
            &\geq f(s_i^{[k+1]}) - \frac{\tau^2}{2\mu}\left\| g^{[k+1]} \right\|^2 + \frac{L}{2}\|\lambda_i^{[k+2]}\|^2\\
            &= f(s_i^{[k+1]}) - \frac{1}{2L}\left\| g^{[k+1]} \right\|^2 + \frac{L}{2}\|\lambda_i^{[k+2]}\|^2. 
            \label{equ:lemma_24_formula_D}
        \end{aligned}
    \end{equation}
    Combining \cite[Lemma~5.7]{lemma_1_support} and \eqref{eq: lip}, we have 
    \begin{equation}
    \begin{aligned}
        f(x_\alpha) \geq& f(x_\beta)
         - \langle \nabla f(x_\alpha), x_\beta-x_\alpha\rangle \\
         &-\frac{L}{2}\|x_\beta-x_\alpha\|^2. \label{equ:lip_another_form}
    \end{aligned}
    \end{equation}
    Substituting $x_\alpha = s_i^{[k+1]}, x_\beta = x_i^{[k+2]}$ into \eqref{equ:lip_another_form}, we can obtain
    \begin{equation}
        \begin{aligned}
            f(s_i^{[k+1]}) &\geq f(x_i^{[k+2]}) - \langle g^{[k+1]}, x_i^{[k+2]} - s_i^{[k+1]} \rangle \\
            &\quad - \frac{L}{2}\left\| s_i^{[k+1]} - x_i^{[k+2]} \right\|^2 \\
            &\overset{\eqref{equ:z_update}}{=} f(x_i^{[k+2]}) - \langle g^{[k+1]},-\frac{g^{[k+1]}}{L} + \lambda_i^{[k+2]} \rangle \\
            &\quad - \frac{L}{2}\left\|-\frac{g^{[k+1]}}{L} + \lambda_i^{[k+2]} \right\|^2 \\
            &= f(x_i^{[k+2]}) + \frac{1}{2L} \|g^{[k+1]}\|^2 - \frac{L}{2} \|\lambda_i^{[k+2]}\|^2. \label{equ:lemma_24_progress_2}
        \end{aligned}
    \end{equation}
    Combining \eqref{equ:lemma_24_formula_D} and \eqref{equ:lemma_24_progress_2}, we have
    \begin{equation}
        \begin{aligned}
            \overset{*}{\Psi_i^{[k+1]}} \geq f(x_i^{[k+2]}). \nonumber
        \end{aligned}
    \end{equation}
    Therefore, by the principle of mathematical induction, $P_k$ holds for every integer $k\geq 1$. \\
    Hence, Lemma~\ref{lemma:lemma_24} is proved. 

\section{Proof of Theorem \ref{theo:convergence_nesterov}} \label{APP: 3}

For convenience, we denote
\begin{subequations}
    \begin{align}
        B_i^{[k]} &:= \Psi_i^{[k]}(x^*) - f(x^*), \label{equ:define_B}\\
        C_i^{[k+1]} &:= \Psi_i^{[k+1]}\left(x^* + \frac{\lambda_i^{[k+2]}}{\tau}\right) - \frac{L}{2}\|\lambda_i^{[k+2]}\|^2 - f(x^*). \label{equ:define_C}
    \end{align}
\end{subequations}
From Lemma~\ref{lemma:lemma_23}, substituting $x=x^*$ into \eqref{equ:lemma_23_main}, we have
\begin{equation}
    \begin{aligned}
        C_i^{[k+1]} &\overset{\eqref{equ:lemma_23_main}}{\leq} (1-\tau) \left(\Psi_i^{[k]}(x^*) - f(x^*)\right) \\
        &\overset{\eqref{equ:define_B}}{=} (1-\tau)B_i^{[k]}. \label{equ:theo_2_formula_G}
    \end{aligned}
\end{equation}
From Lemma~\ref{lemma:lemma_21}, we have 
\begin{equation}
    \begin{aligned}
        C_i^{[k+1]} &= \overset{*}{\Psi}_i^{[k+1]} + \frac{\mu}{2} \left\| x^* + \frac{\lambda_i^{[k+2]}}{\tau} - \phi_i^{[k+1]} \right\|^2 \\
        &\quad - \frac{L}{2}\|\lambda_i^{[k+2]}\|^2 - f(x^*). 
        \label{equ:theo_2_formula_E}
    \end{aligned}
\end{equation}
Substituting $x_\alpha=s_i^{[k+1]}, x_\beta=s_i^{[k+1]}-\frac{1}{L}g^{[k+1]}$ into \eqref{equ:lip_another_form}, we can get
\begin{equation}
    \begin{aligned}
        f\left( s_i^{[k+1]}-\frac{1}{L}g^{[k+1]} \right) & \leq f(s_i^{[k+1]}) - \langle g^{[k+1]}, \frac{g^{[k+1]}}{L} \rangle\\
        &\quad + \frac{1}{2L}\|g^{[k+1]}\|^2 \\
        &= f(s_i^{[k+1]}) - \frac{1}{2L}\|g^{[k+1]}\|^2\\
        &\overset{\eqref{equ:lemma_24_formula_D}}{\leq} \overset{*}{\Psi}_i^{[k+1]} - \frac{L}{2} \|\lambda_i^{[k+2]}\|^2. \label{equ:theo_2_progress_1}
    \end{aligned}
\end{equation}
Combining \eqref{equ:theo_2_progress_1} with $f(x^*)\leq f(x)$ for any $x \in \mathbb{R}^p$, we have
\begin{equation}
    \begin{aligned}
        \overset{*}{\Psi}_i^{[k+1]} - \frac{L}{2} \|\lambda_i^{[k+2]}\|^2 - f(x^*) \geq 0. 
        \label{equ:theo_2_progress_2}
    \end{aligned}
\end{equation}
So we have
\begin{equation}
    \begin{aligned}
        C_i^{[k+1]} \overset{\eqref{equ:theo_2_formula_E}}{\geq} \frac{\mu}{2} \left\| x^* + \frac{\lambda_i^{[k+2]}}{\tau} - \phi_i^{[k+1]} \right\|^2\geq 0. 
        \label{equ:theo_2_formula_F}
    \end{aligned}
\end{equation}
For convenience, we denote
\begin{align}
    \tilde{\phi}_i^{[k+1]} := \phi_i^{[k+1]} - \frac{\lambda_i^{[k+2]}}{\tau}. \label{equ:define_tilde_phi}
\end{align}
By the Cauchy–Schwarz inequality, we have
\begin{equation}
    \begin{aligned}
        &\quad\frac{\mu}{\tau}\left| \langle \tilde{\phi}_i^{[k+1]}-x^*, \lambda_i^{[k+2]} \rangle \right| \\
        &\leq \frac{\mu}{\tau} \left\| \tilde{\phi}_i^{[k+1]}-x^* \right\|\cdot\|\lambda_i^{[k+2]}\| \\
        &\overset{\eqref{equ:theo_2_formula_F}}{\leq} \frac{\mu}{\tau}\sqrt{\frac{2C_i^{[k+1]}}{\mu}} \|\lambda_i^{[k+2]}\|\\
        &\leq 2\sqrt{LC_i^{[k+1]}} \|\lambda_i^{[k+2]}\|. 
        \label{equ:theo_2_cs}
    \end{aligned}
\end{equation}
Combining Lemma~\ref{lemma:lemma_21}, we can obtain
\begin{equation}
    \begin{aligned}
        B_i^{[k+1]} &= \overset{*}{\Psi}_i^{[k+1]} - f(x^*) + \frac{\mu}{2}\left\| x^* - \phi_i^{[k+1]}\right\|^2 \\
        &\overset{\eqref{equ:theo_2_formula_E}}{=} C_i^{[k+1]} + \frac{L}{2}\|\lambda_i^{[k+2]}\|^2 + \frac{\mu}{2}\Bigg( \left\|x^* - \phi_i^{[k+1]}\right\|^2 \\
        &\quad - \left\|x^* - \phi_i^{[k+1]} + \frac{\lambda_i^{[k+2]}}{\tau} \right\|^2 \Bigg) \\
        &= C_i^{[k+1]} + \frac{\mu}{\tau}\langle \tilde{\phi}_i^{[k+1]} - x^*, \lambda_i^{[k+2]} \rangle + L\|\lambda_i^{[k+2]}\|^2 \\
        &\overset{\eqref{equ:theo_2_cs}}{\leq} C_i^{[k+1]} + 2\sqrt{LC_i^{[k+1]}} \|\lambda_i^{[k+2]}\| + L\|\lambda_i^{[k+2]}\|^2 \\
        &= \left( \sqrt{C_i^{[k+1]}} + \sqrt{L} \|\lambda_i^{[k+2]}\| \right)^2\\
        &\overset{\eqref{equ:theo_2_formula_G}}{\leq} \left( \sqrt{1-\tau}\cdot \sqrt{B_i^{[k]}} + \sqrt{L} \|\lambda_i^{[k+2]}\| \right)^2. 
        \label{equ:theo_2_progress_3}
    \end{aligned}
\end{equation}
So from \eqref{equ:theo_2_progress_3}, we have
\begin{equation}
    \begin{aligned}
        \sqrt{B_i^{[k+1]}} &\leq \sqrt{1-\tau}\cdot \sqrt{B_i^{[k]}} + \sqrt{L} \|\lambda_i^{[k+2]}\| \\
        &\leq \left( \sqrt{1-\tau} \right)^k\cdot \sqrt{B_i^{[1]}} + \sqrt{L}\cdot\sum_{j=1}^k \left( \sqrt{1-\tau} \right)^{k-j} \\
        &\quad \cdot\|\lambda_i^{[j+2]}\| \\
        &\overset{\eqref{equ:lemma_1_main}}{\leq} \left( \sqrt{1-\tau} \right)^k \sqrt{B_i^{[1]}} + 2\sqrt{Lp}\Delta \sum_{l=0}^{k-1}\left( \sqrt{1-\tau} \right)^l \\
        &\leq \left( \sqrt{1-\tau} \right)^k \sqrt{B_i^{[1]}} + \frac{2\sqrt{Lp}\Delta}{1-\sqrt{1-\tau}}. 
        \label{equ:theo_2_progress_4}
    \end{aligned}
\end{equation}
Combining \eqref{equ:theo_2_progress_4} with Lemma~\ref{lemma:lemma_24}, we have
\begin{equation}
    \begin{aligned}
        &\quad\quad f(x_i^{[k+1]}) - f(x^*) \\
        &\overset{\eqref{equ:lemma_24_main}}{\leq} \overset{*}{\Psi}_i^{[k]} - f(x^*) \\
        &\overset{\eqref{equ:define_B}}{\leq} B_i^{[k]} \\
        &\overset{\eqref{equ:theo_2_progress_4}}{\leq} \left( \left( \sqrt{1-\tau} \right)^{k-1} \sqrt{B_i^{[1]}} + \frac{2\sqrt{Lp}\Delta}{1-\sqrt{1-\tau}}\right)^2 \\
        &\leq 2(1-\tau)^{k-1} B_i^{[1]} + \frac{8Lp\Delta^2}{\left( 1-\sqrt{1-\tau} \right)^2}.
    \end{aligned}
\end{equation}
Therefore, Theorem~\ref{theo:convergence_nesterov} is then proved. 

\bibliographystyle{IEEEtran}  
\bibliography{references}      

\end{document}